\documentclass{PoS}

\title{Charm physics with highly improved staggered quarks}

\ShortTitle{Charm physics with highly improved staggered quarks}

\author{HPQCD and UKQCD:
C.T.H. Davies,$^a$  \speaker{E. Follana},$^a$
K. Hornbostel,$^b$ G.P. Lepage,$^c$ Q. Mason,$^d$ H. Trottier,$^e$ 
J. Shigemitsu$^f$ and K. Wong$^a$ \\
\llap{$^a$} University of Glasgow, Glasgow, UK \\
\llap{$^b$} Dallas Southern Methodist University, Dallas,  Texas, USA\\ 
\llap{$^c$} Cornell University, Ithaca, New York, USA \\
\llap{$^d$} Cambridge University, Cambridge, UK \\
\llap{$^e$} Simon Fraser University, Vancouver, British Columbia, Canada \\
\llap{$^f$} Ohio State University, Columbus, Ohio, USA }

\abstract{ We use a relativistic highly improved staggered quark
action to discretize charm quarks on the lattice. We calculate the
masses and the dispersion relation for heavy-heavy and heavy-light
meson states, and show that for lattice spacings below .1 fm, the
discretization errors are at the few percent level. We also discuss
the prospects for accurate calculations at the few percent level of
$f_{D_s}$, $f_D$, and the leptonic width of the $\psi$ and $\phi$. }

\FullConference{XXIVth International Symposium on Lattice Field Theory\\
                July 23-28, 2006\\
                Tucson, Arizona, USA}

\begin{document}

\section{Introduction}

\begin{figure}
\begin{center}
\includegraphics[width=3in]{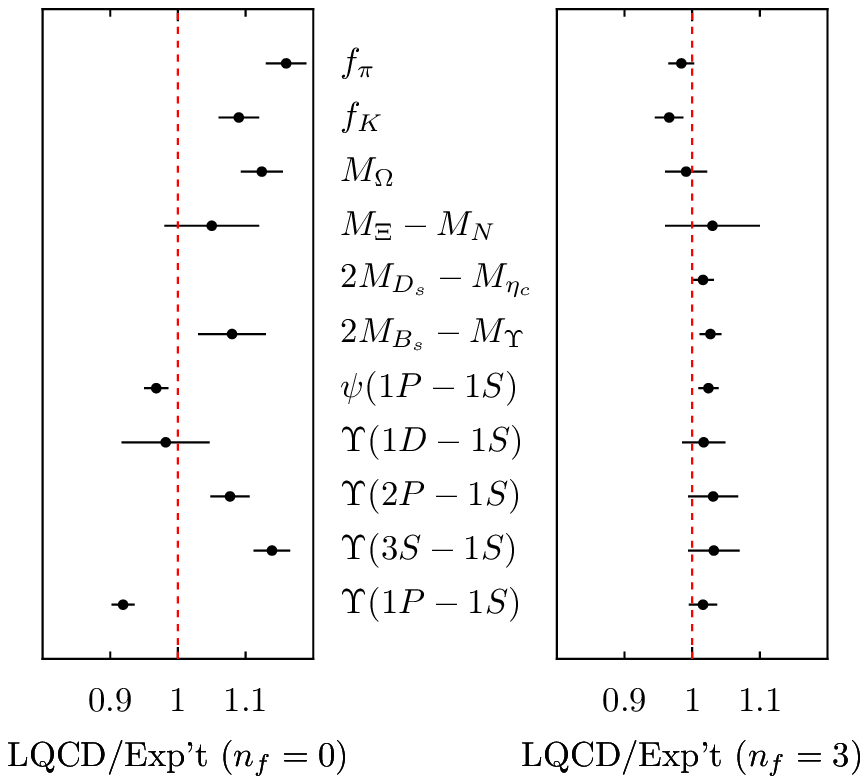}
\end{center}
\caption{\label{ratio} Comparison between quenched and unquenched
simulation results.}
\end{figure}

Improved staggered quarks have proved very effective in obtaining
precise results of phenomenological interest in the unquenched light
valence sector \cite{PRL1} (see figure \ref{ratio} for a comparison
of quenched and unquenched results.) On the other hand,
non-relativistic effective field formulations are very successful in
the bottom sector \cite{NRQCD1, NRQCD2}.  Although a non-relativistic
formulation can also be applied, in principle, to the charm sector,
the errors are much larger.

Highly improved staggered quarks have very small discretization
errors. Combining this with fine enough lattices may provide a good
method of handling charm quarks.

CLEO-c is making precise measurements of several quantities (for
example $f_{D_s}$ $f_D$ ) in the charm system with small errors
($\approx 4 \%$).  For comparison, Fermilab results have errors of
$\approx 8 \%$ \cite{FNAL}.  This provides a good opportunity to test
our methods.

\section{Improved Staggered Quarks}

The massless one-link (Kogut-Susskind) staggered Dirac operator is
defined as:

\begin{equation}
D (x,y)  =  \frac{1}{2au_0} \sum_{\mu=1}^d
\eta_\mu(x) \left[ U_\mu(x) \delta_{x+\hat{\mu},y} - H.c. \right] \; , \quad
\eta_\nu(x)  =  (-1)^{\sum_{\mu < \nu} x_\mu }
\end{equation}

with $u_0$ an optional tadpole-improvement factor. 

This operator suffers from doubling: there are four ``tastes''
(non-physical flavours) of fermions in the spectrum, which couple
through taste-changing interactions. These are lattice artifacts of
order $a^2$, involving at leading order the exchange of a gluon of
momentum $q \approx \pi/a$.  Such interactions are perturbative for
typical values of the lattice spacing, and can be corrected
systematically a la Symanzik. By judiciously smearing the gauge field
we can remove the coupling between quarks and high momentum gluons.

\begin{figure}
\center{
  \includegraphics[width=3in]{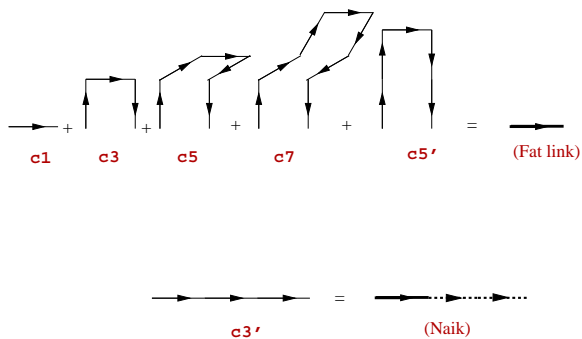}
}
\caption{\label{asqtad}
Paths used for smearing in the ASQTAD staggered action.
}
\end{figure}

The most widely used improved staggered action is called ASQTAD, and
removes all tree-level $a^2$ discretization errors \cite{Naik,
Lepage1, Orginos}. The paths used to smear the gauge-fields are shown
in figure \ref{asqtad}.

The HISQ (highly improved staggered quarks) staggered Dirac operator
involves two levels of smearing with an intermediate projection onto
$SU(3)$. It is designed so that, as well as eliminating all $a^2$
discretization errors, it further reduces the one-loop taste-changing
errors (for a more detailed discussion see \cite{HISQ}.) 

This action has been shown to substantially reduce the errors
associated with the taste-changing interactions \cite{HISQ, spectrum1,
spectrum2}. 

\section{Charm Sector}

When we put massive quarks on the lattice, the discretization errors
grow with the quark mass as powers of $a m$. Therefore to obtain small
errors we would need $a m \ll 1$. For heavy quarks this would require
very small lattice spacings. On the other hand, to keep our lattice
big enough to accommodate the light degrees of freedom, we need $L a
\gg m_\pi^{-1}$. The fact that we have two very different scales in
the problem makes difficult a direct solution. What we can do instead
is to take advantage of the fact that $m$ is large, by using an
effective field theory (NRQCD, HQET). This program has been very
successful for b quarks \cite{NRQCD1, NRQCD2, FNAL}.

The charm quark is in between the light and heavy mass regime. It is
quite light for an easy application of NRQCD, but quite large for the
usual relativistic quark actions, $a m_c \stackrel {\textstyle
<}{\sim} 1$. However, if we use a very accurate action (HISQ) and fine
enough lattices (fine MILC ensembles), it is possible to get results
accurate at the few percent level.


\section{Results}

We use $2+1$-flavours unquenched configurations generated by the MILC
collaboration \cite{MILC1, MILC2, MILC3}. We present here results
obtained from an ensemble with $m_l = m_s / 5$, where $m_s$ is the
light and $m_s$ the strange quark mass, and with a lattice spacings of
$1/11 fm$. The extent of the corresponding lattices is $28^3 x \; 96$.

The bare mass used for the charm valence quarks is fixed by adjusting
the ``Goldstone'' pseudoscalar mass to the experimental value.


\subsection{$c\bar c$ pseudoscalar and vector}

\begin{figure}
\begin{center}
\includegraphics[width = 0.6 \hsize, angle=-90]{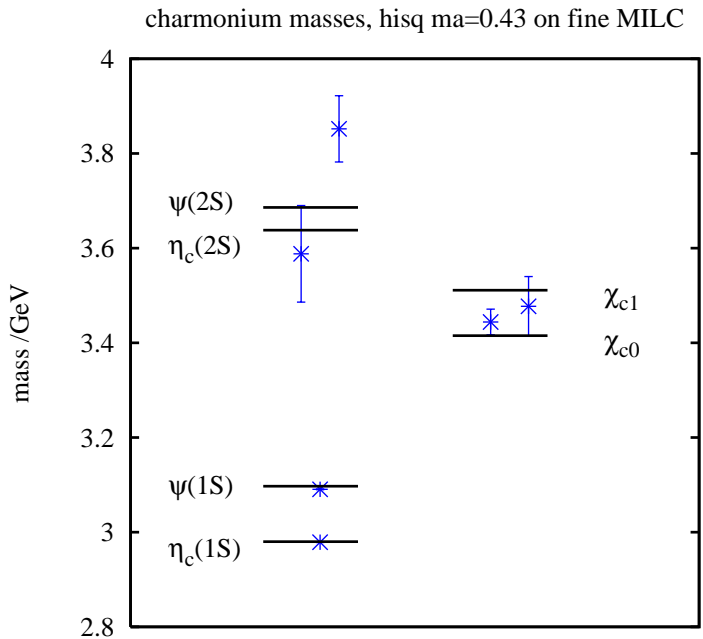}
\end{center}
\caption{\label{spectrum} Charmonium spectrum on fine MILC configurations.}
\end{figure}

We show in figure \ref{spectrum} some of our results for the
charmonium spectrum, as a check of the formalism. We haven't optimized
in any way our operators for the calculation of the excited states,
which therefore have large errors.

\begin{figure}
\begin{center}
\includegraphics[width = .6 \hsize, angle=-90]{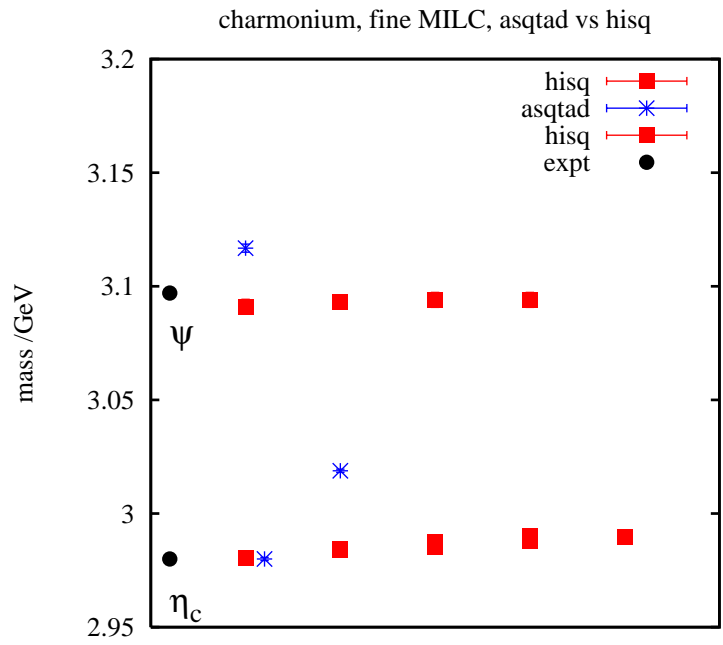}
\end{center}
\caption{\label{spacing} Charm pseudoscalar and vector states for 
HISQ and ASQTAD on fine MILC configurations.}
\end{figure}

In figure \ref{spacing} we show the results for the pseudoscalar and
vector mesons. In this sectors we have very small statistical
errors. In the staggered formalism we have 16 different mesons of each
type, with, in general, different masses, due to the taste-changing
interactions. We can see that such mass splitting is mostly noticeable
for the pseudoscalar states, but is almost negligible for the vector
mesons, where it's below our statistical error. This is consistent
with previous results \cite{JLQCD}, although the actual size of the
splitting is very much reduced with respect to the one using one-link
staggered fermions. The total splitting for $\eta_c$ is only about 10
Mev for HISQ, and is also much smaller than the one for ASQTAD. We
obtain an hyperfine splitting of $\approx 110 MeV$, to be compared
with the experimental value of $117 MeV$.

\subsection{Speed of Light}

\begin{figure}
\begin{center}
\includegraphics[width= .6 \hsize,angle=-90]{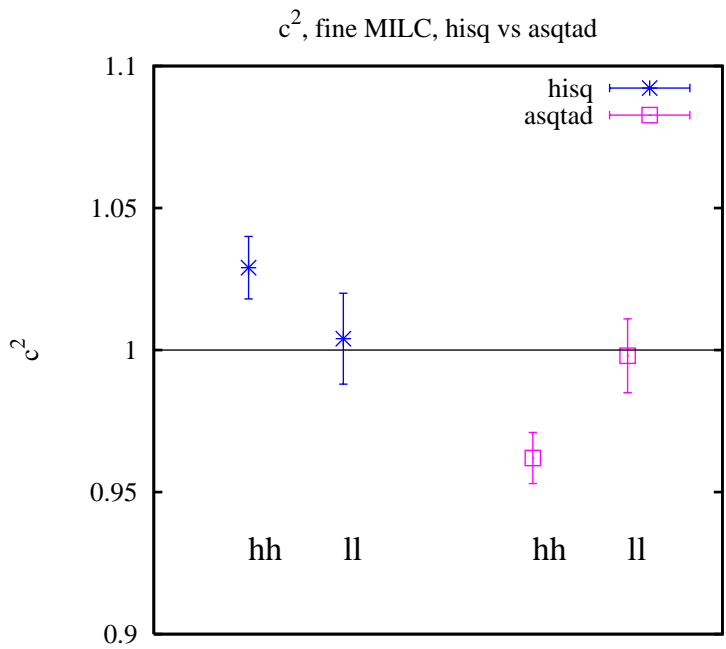}
\end{center}
\caption{\label{c2} Speed of light squared for HISQ and ASQTAD on 
fine MILC configurations.}
\end{figure}

It is important to check that the discretization artifacts at the
masses we use in our simulation do not spoil the relativistic
invariance of the action (as is bound to happen at large enough mass.)
One way to do this is by calculating the dispersion relation of a
meson, or equivalently, that the squared speed of light is still 1 at
small non-zero momenta.

In figure \ref{c2} we show the results for $c^2$ at a small non-zero
momentum for pseudoscalar heavy-heavy and light-light mesons, where
the heavy mass is set to the charm mass and the light one to the
strange mass. We show the results for both HISQ and ASQTAD
actions. $c^2$ is one at the strange mass, but different from 1 at the
charm mass, and that deviation is larger for the ASQTAD action than
for the HISQ action. 

\begin{figure}
\begin{center}
\includegraphics[width=.6 \hsize,angle=-90]{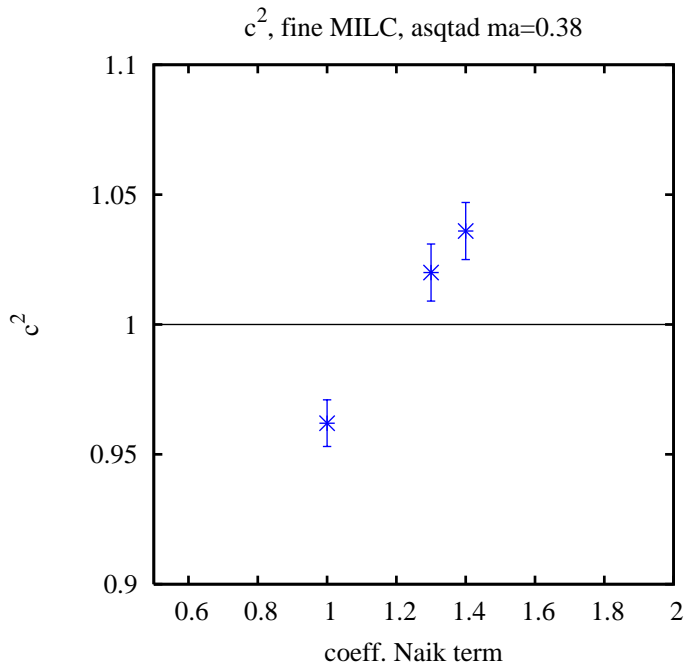}
\end{center}
\caption{\label{c2naik} Speed of light squared as a function of the coefficient 
of the Naik term for ASQTAD on fine MILC configurations.}
\end{figure}

This error can be corrected for by modifying the overall coefficient
of the Naik term in the action. It can be shown that this simple
modification to the action removes all discretization errors of order
$\alpha_s (a m)^2$, the largest ones remaining in our improved actions
for large quark mass \cite{HISQ}. We plot in figure \ref{c2naik} the
result of modifying such coefficient in $c^2$ for the ASQTAD action.

\section{Conclusions and Outlook}

  The use of a highly improved quark action and fine enough lattices
provides a very good way of studying the charmonium systems from first
principles. We have shown that the discretization errors are well
under control, and we can obtain precise results in the charm sector.

We are now using this formalism for the precision calculation of
several interesting quantities, which can be checked against the new
CLEO-c results. This includes $f_{D_s}$ and $f_D$. We could also use
this method to obtain leptonic decay widths $D \rightarrow \mu
\nu_\mu$.

\section{Acknowledgments}

This work is supported by PPARC, NSF and DOE.

The calculations were carried out on computer clusters at Scotgrid and
QCDOCX. QCDOCX is funded by PPARC JIF grant PPA/J/S/1998/00756. 
We thank David Martin and EPCC for assistance.

\end{document}